\documentstyle[12pt,psfig]{article}
\def\be{\begin{equation}}
\def\ee{\end{equation}}
\def\bear{\be\begin{array}}
\def\eear{\end{array}\ee}
\def\bea{\begin{eqnarray}}
\def\eea{\end{eqnarray}}

\def\baselinestretch{1}
\begin{document}
%%%%%%%%%%%%%%%%%%%%%%%%%%% subequations.sty %%%%%%%%%%%%%%%%%%%%%%%%
\catcode`@=11
\newtoks\@stequation
\def\subequations{\refstepcounter{equation}%
  \edef\@savedequation{\the\c@equation}%
  \@stequation=\expandafter{\theequation}%   %only want \theequation
  \edef\@savedtheequation{\the\@stequation}% % expanded once
  \edef\oldtheequation{\theequation}%
  \setcounter{equation}{0}%
  \def\theequation{\oldtheequation\alph{equation}}}
\def\endsubequations{\setcounter{equation}{\@savedequation}%
  \@stequation=\expandafter{\@savedtheequation}%
  \edef\theequation{\the\@stequation}\global\@ignoretrue

\noindent}
\catcode`@=12
%%%%%%%%%%%%%%%%%%%%%%%%%%%%%%%%%%%%%%%%%%%%%%%%%%%%%%%%%%%%%%%%%%%%%
\begin{titlepage}
\title{{\bf Constraining Supergravity Scenarios through the
$b\rightarrow s,\gamma$ Decay}\thanks{Work partly supported
by CICYT under contract AEN94-0928, and  by the European Union under
contract
No. CHRX-CT92-0004.}
}
\vspace{0.9cm}
\author{{\bf B. de Carlos}\thanks{
Supported by a Spanish M.E.C. Postdoctoral Fellowship.} $\;\;$ and
$\;\;$ {\bf
J.A. Casas}\thanks{On leave of absence from Instituto de Estructura de
la Materia (CSIC), Serrano 123, 28006-Madrid, Spain.} \\
Department of Theoretical Physics, University of Oxford \\
1 Keble Road, Oxford OX1 3NP, UK}
\vspace{0.5cm}

\date{}
\maketitle
%\vspace{1.5cm}
\def\baselinestretch{1.15}
\begin{abstract}
\noindent
We evaluate the branching ratio BR($b\rightarrow s,\gamma$) in the
minimal supersymmetric standard model (MSSM), determining the
corresponding phenomenological restrictions on two attractive
supergravity scenarios, namely minimal supergravity and a class of
models with a natural solution to the $\mu$ problem. We have included
in the calculation some one--loop refinements that have a substantial
impact on the results. The numerical results show some disagreements
with part of the previous results in the literature, while they are in
agreement with others. For minimal supergravity the CLEO upper and lower
bounds put important restrictions on the scalar and gaugino masses in
both cases $\mu<0$ and $\mu>0$. For the other supergravity scenarios
the relevant CLEO bound is the upper one. It is stressed the fact that
an eventual improvement of the experimental bounds of order $10^{-4}$
would strengthen the restrictions on the MSSM dramatically. This would
be enough to discard these supergravity scenarios with $\mu<0$ if
no discrepancy is found with the standard model prediction, while for
$\mu>0$ there will remain low-energy windows.

\end{abstract}

\thispagestyle{empty}

\leftline{}
\leftline{January 1995}
\leftline{}

\vskip-19.5cm
\rightline{}
\rightline{{\bf OUTP--95--01P}}
\rightline{{\bf IEM--FT--99/95}}

\vskip3in

\end{titlepage}
\newpage
%%%%%%%%%%%%%%%%%%%%%%%%%%%%%%%%%%%%%%%%%%%%%%%%%%%%%%%%%%%%%%%%%%%
\setcounter{page}{1}

\section{Introduction}

It is well known that the $b\rightarrow s,\gamma$ process has the
potential to put relevant constraints to physics beyond the Standard
Model (SM). This comes mainly from two facts. First, as a FCNC
process, the $b\rightarrow s,\gamma$ decay occurs only beyond the tree
level, thus being sensitive to the characteristics of possible new
particles circulating in the relevant loops. Second, there are
available experimental data on the exclusive \cite{ammar93}
$B\rightarrow K^*, \gamma$ and the inclusive \cite{browd94}
$B\rightarrow X_s, \gamma$ decays that lead to upper and lower bounds
on the branching ratio BR($b\rightarrow s,\gamma$) of the same order
as the SM prediction. In particular, from the inclusive $B$ decay:
$1\times 10^{-4}<$ BR($b\rightarrow s,\gamma$) $<4\times 10^{-4}$ .
There has recently been a large amount of work discussing the
prediction for BR($b\rightarrow s,\gamma$) in the SM and extensions
of it, particularly in the Minimal Supersymmetric Standard Model (MSSM)
\cite{berto91,hewet93,marco94,berto94,barge94,goto94}. The latter scenario
probably represents the most promising and theoretically well founded
extension of the SM capable of giving new measurable physics
\cite{haber85,rosslibr,baili94}. In it, besides the usual SM diagram
with a $W$ gauge boson and a top quark in the loop, there are
additional contributions coming from loops involving charged Higgses
($H^{-}$) and a top quark, charginos ($\chi^{-}$) and u-type squarks
(of which the relevant contributions come from the stops,
$\tilde{t}_{L,R}$, and scharms, $\tilde{c}$), and a gluino
($\chi_{3}$) or neutralinos ($\chi_{i}^{0}$) plus a d-type squark
(mainly $\tilde{b}$ and $\tilde{s}$) \cite{berto91}. As
pointed out in ref.~\cite{berto91}, the latter two diagrams do not
contribute significantly to the BR and can therefore be neglected.
So we end up with the following expression for the branching ratio
at next to leading order (in units of the BR for the semileptonic $b$
decay):
\begin{eqnarray}
\frac{BR(b \rightarrow s \gamma)}{BR(b \rightarrow c e \bar{\nu})} & = &
\frac{|K_{ts}^{*} K_{tb}|^{2}}{|K_{cb}|^{2}} \frac{6 \alpha_{QED}}{\pi}
\nonumber \\
& \times & \frac{\left [\eta^{16/23} A_{\gamma} + \frac{8}{3} (\eta^{14/23} -
\eta^{16/23}) A_{g} + C \right]^{2}}{I(z)} F
\label{BR}
\end{eqnarray}
where $z=\frac{m_c}{m_b}$, $\eta=\frac{\alpha_s(M_W)}{\alpha_s(m_b)}$,
$I(z)= 1-8z^{2}+8z^{6}-z^{8}-24z^{4}\log(z)$ is the phase space factor,
$C$ stands for the leading logarithmic QCD corrections which have been
calculated in refs.~\cite{berto87,grins90,buras93,buras94}, and
$F \sim 1-\frac{8}{3}
\frac{\alpha_s(m_b)}{\pi} \frac{1}{\kappa(z)}$ contains NLO effects
($\kappa(z)$ being the NLO correction to the semileptonic
decay)\footnote{Notice
that we have chosen $Q=m_{b}$ as our renormalization scale. For a
discussion about the uncertainties derived from this election see
ref.~\cite{buras94}.}. Finally,
$A_{\gamma,g}$ are the coefficients of the effective operators for the
$bs\gamma$ and $bsg$ interactions; in our case, as mentioned before, we
consider as relevant the contributions coming from the SM diagram plus
those with top quark and charged Higgs, and stops/scharms and
charginos running in the loop. Their expressions are given by:
\begin{eqnarray}
A_{\gamma,g}^{SM} & = & \frac{3}{2} \frac{m_t^{2}}{M_W^{2}}
f_{\gamma,g}^{(1)} \left( \frac{m_t^{2}}{M_W^{2}} \right) \nonumber \\
A_{\gamma,g}^{H^{-}} & = & \frac{1}{2} \frac{m_t^{2}}{m_H^{2}} \left[
\frac{1}{\tan^{2} \beta} f_{\gamma,g}^{(1)} \left(
\frac{m_t^{2}}{m_H^{2}} \right)
+ f_{\gamma,g}^{(2)} \left( \frac{m_t^{2}}{m_{H}^{2}} \right) \right]
\nonumber \\
A_{\gamma,g}^{\chi^{-}} & = & \sum_{j=1}^{2} \left\{
\frac{M_W^{2}}{M_{\chi_j}^{2}} \left[ |V_{j1}|^{2} f_{\gamma,g}^{(1)}
\left( \frac{m_{\tilde{c}}^{2}}{M_{\chi_j}^{2}} \right) - \sum_{k=1}^{2}
\left| V_{j1} T_{k1} - \frac{V_{j2} m_t T_{k2}}{\sqrt{2} M_W \sin\beta}
\right|^{2} \right. \right. \nonumber \\
& \times & \left. \left.
f_{\gamma,g}^{(1)} \left( \frac{m_{\tilde{t}_k}^{2}}{M_{\chi_j}^{2}} \right)
\right] - \frac{U_{j2}}{\sqrt{2} \cos \beta} \frac{M_W}{M_{\chi_j}} \left[
V_{j1} f_{\gamma,g}^{(3)} \left( \frac{m_{\tilde{c}}^{2}}{M_{\chi_j}^{2}}
\right)
\right. \right. \label{amps} \\
& - & \left. \left. \sum_{k=1}^{2} \left( V_{j1} T_{k1} - V_{j2}
T_{k2} \frac{m_t}
{\sqrt{2} M_W \sin \beta} \right) T_{k1} f_{\gamma,g}^{(3)} \left(
\frac{m_{\tilde{t}_k}^{2}}{M_{\chi_j}^{2}} \right) \right] \right\}
\nonumber
\end{eqnarray}
where the functions $f_{\gamma,g}^{(i)}$, $i=1,2,3$ were originally
calculated in ref.~\cite{berto91}, and all the masses are understood
to be at the electroweak scale. $V$ and $U$ are the matrices which
diagonalise the chargino mass matrix, while $T$ diagonalises the stop
mass matrix.

As we can see from the former expressions, it is not easy to draw
general conclusions about the behaviour of these amplitudes due to
their complicated dependence on the different masses and matrix
elements. However, there are some general features which are worth
mentioning: for example, both the SM and the charged Higgs
contributions have always the same sign, giving therefore a total
amplitude which is bigger than the SM one. This fact has been used by
several authors \cite{grins90,rizzo88} to impose severe restrictions
on two Higgs doublets models (2HDM); in our case this effect is not as
sharp due to the presence of the chargino contribution which, for a
wide range of the parameter space, has opposite sign to that of the
other two amplitudes. As we will see this is not enough, in general,
to lower the total BR below the SM prediction, although in some cases
it might be possible to have such big values for
$A_{\gamma,g}^{\chi^{-}}$ as to drive the total BR even below the
CLEO lower bound.

In the formulation of the MSSM the mass matrices of charginos, charged
Higgses, stops, etc, appearing in the above expressions, are not
independent parameters. They come as a low energy result of the actual
initial parameters of the theory, which are assumed to be given at the
unification scale, $M_X$. It is therefore of great importance to
correctly establish  the constraints on these parameters from
$b\rightarrow s,\gamma$ measurements. Several recent papers have been
concerned with this task \cite{hewet93,marco94,berto94,barge94,goto94}.
However, as we will see, the complete one--loop effects to be taken
under consideration have not been always properly included,
particularly in the works dealing with the minimal supergravity
(SUGRA) scenario. A careful inclusion of these effects, especially the
radiative electroweak breaking, is precisely the main goal of this
article.

In addition, most of the SUGRA analysis have just dealt with the minimal
SUGRA scenario, which is only justified by its simplicity. In fact,
this scenario has a naturalness problem, the so--called
$\mu$--problem, namely the origin of the electroweak size of the
$\mu$ parameter in the superpotential (see below). In recent times,
however, there have appeared string and grand unification well
motivated mechanisms to solve the $\mu$--problem
\cite{giuma,cmmu,giurou,kaplu93,bim} that imply a departure from
minimal SUGRA. It is also the aim of this article to study the
constraints on these scenarios coming from the $b\rightarrow s,\gamma$
process.

\vspace{0.3cm}
\noindent
In section 2 we introduce two supergravity scenarios, labelled SUGRA I
and SUGRA II, which correspond to the previously mentioned supergravity
frameworks, and establish their connection to the MSSM. In section 3
we explain our approach to determine the restrictions on the MSSM
coming from the $b\rightarrow s,\gamma$ decay, focussing our attention on
some one--loop refinements that have a substantial impact on the results.
The numerical results for the supergravity scenarios under consideration
and their implications are presented and discussed in section 4. Finally,
we present our conclusions in section 5.

\section{MSSM and SUGRA scenarios}

The MSSM is defined by the superpotential, $W$, and the form of the
soft supersymmetry breaking terms. $W$ is given by
\be
\label{W}
W=\sum_i\left\{
h_{u_i}Q_i H_2 u_i^c + h_{d_i}Q_i H_1 d_i^c
+ h_{e_i}L_i H_1 e_i^c \right\} +  \mu H_1 H_2\;\; ,
\ee
where $i$ is a generation index, $Q_i$ ($L_i$) are the scalar partners
of the quark (lepton) SU(2) doublets, $u^c_i,d^c_i$ ($e^c_i$) are the
quark (lepton) singlets and $H_{1,2}$ are the two SUSY Higgs doublets;
the $h$--factors are the Yukawa couplings and $\mu$ is the usual Higgs
mixing parameter. In all terms of eq.~(\ref{W}) the usual SU(2)
contraction is assumed, e.g. $\mu\epsilon_{ij} H_1^i H_2^j$ with
$\epsilon_{12}=-\epsilon_{21}=1$. From $W$ the
(global) supersymmetric part of the Lagrangian is
readily obtained
\be
\label{Lsusy}
{\cal L}_{SUSY}=-\sum_j\left|\frac{\partial W}{\partial
\phi_j}\right|^2
-\frac{1}{2}\left(\sum_{j,k}\left[\frac{\partial^2 W}{\partial
\phi_j \partial \phi_k} \right] + {\rm h.c.}\right)\ +\ {\rm D-terms}\;\;,
\ee
where $\phi_{j,k}$ ($\psi_{j,k}$) run over all the scalar (fermionic)
components of the chiral superfields.
In addition to this, the soft breaking terms coming
from the (unknown) supersymmetry breaking mechanism have the form
\bea
\label{Lsoft}
{\cal L}_{\rm soft}&=&\frac{1}{2}M\lambda_a\lambda_a -\left\{\;\;
\sum_j m^2 |\phi_j|^2\ +\ \sum_i A\left[
h_{u_i}Q_i H_2 u_i^c + h_{d_i}Q_i H_1 d_i^c
\right.\right.
\nonumber \\
&+& \left. \left. h_{e_i}L_i H_1 e_i^c + {\rm h.c.} \right]
+ \left[ B\mu H_1 H_2 + {\rm h.c.}\right]\phantom{\sum_{j}}
\right\} \;\; ,
\eea
where $a$ is gauge group index, $\lambda_a$ are the gauginos,
and the remaining fields in the formula denote just the corresponding
scalar components. The parameters $m$, $M$, $A$, $B$ are the
universal scalar and
gaugino masses and the universal coefficients of the trilinear and
bilinear scalar terms, respectively. The universality is assumed at
the unification scale $M_X$.

With the previous definitions the chargino mass term in matrix form,
which plays a crucial role in the expresions for
BR($b\rightarrow s,\gamma$) [see eqs.~(\ref{BR}, \ref{amps})], is given
by
\bea
\label{Mchar}
-\ (-i\lambda^{-},\tilde{H}_1^{-})
\left(
\begin{array}{cc}
M_2 & \sqrt{2}M_W\sin\beta \\
\sqrt{2}M_W\cos\beta & -\mu
\end{array}
\right)
\left(
\begin{array}{c}
-i\lambda^{+} \\
\tilde{H}_2^{+}
\end{array}
\right)\;\;,
\eea
where $M_2$ is the (renormalized) SU(2) gaugino mass, $\lambda^{\pm}=
(\lambda_1\mp i \lambda_2)/\sqrt{2}$ and the tildes denote fermionic
components.

As has been noted in ref.~\cite{berto94}, the sign of $\mu$ in the
chargino mass matrix has been incorrectly written in
ref.~\cite{gunio86}; this is also the case for the other diagonal
entry ($M_2$). The same mistake with the $\mu$ sign occurs in
ref.~\cite{baili94}. On the other hand, the signs of the chargino mass
matrix have been consistently written
in refs.~\cite{haber85,ibane85,barge94,berto94}.

\vspace{0.2cm}
\noindent
The particular values of the soft breaking parameters $m,M,A,B$
depend on the type of SUGRA theory from which the MSSM derives and on
the mechanism of supersymmetry breaking. We will consider two types
of SUGRA theories without specifying the mechanism of supersymmetry
breaking.

\vspace{0.3cm}
\noindent {\bf SUGRA I}

\vspace{0.2cm}
\noindent This is just the minimal supergravity theory. It is defined
by a K\"ahler potential $K=\sum_j|\phi_j|^2$ and a gauge kinetic
function $f_{ab}=\delta_{ab}$, so that all the kinetic terms are
canonical, whereas the superpotential $W$ is assumed to be as in
eq.~(\ref{W}), $\mu$ being an inicial parameter. Then, irrespectively
of the supersymmetry breaking mechanism, the gaugino and scalar
masses are automatically universal and the coefficients $A$ and $B$
are universal and related to each other by the well known relation
\be
\label{ABmin}
B=A-m \;\; .
\ee
This supergravity theory is attractive for its simplicity and for the
natural explanation that offers to the universality of the soft
breaking terms. Actually, universality is a desirable property not
only to reduce the
number of independent parameters, but also for phenomenological
reasons, particularly to avoid unwanted FCNC and CP violating effects
(see e.g. ref.~\cite{rosslibr}). However, this scenario has a serious
drawback,
namely the unnaturally small (electroweak) size of the initial $\mu$
parameter in the superpotential. This is the so-called $\mu$ problem,
that leads us to the next supegravity scenario.

\vspace{0.3cm}
\noindent {\bf SUGRA II}

\vspace{0.2cm}
\noindent Recently, there have appeared several attractive mechanisms
to solve the $\mu$ problem that, quite remarkably,
lead to a
similar prediction for the value of $B$. Very briefly, these are the
following:

\begin{description}

\item[{\em a)}]

\hspace{10mm} In ref.~\cite{cmmu} was noted that if the superpotential
has a non-renormalizable term of the form
\be
\label{lamWo}
\lambda W_o H_1 H_2\;\;,
\ee
where $W_o$ is the renormalizable superpotential and $\lambda$ an
unknown coupling, then a $\mu$ term is automatically generated with
size $\mu=\lambda m_{3/2}$. The $B$ parameter can also be
straightforwardly evaluated. Assuming that the K\"ahler potential is
as in minimal supergravity in order to guarantee universal soft
breaking terms, the simple result is
\be
\label{ABstr}
B=2m\;\;.
\ee
For this mechanism to work, $\mu$ must be vanishing at the
renormalizable level, a fact that, remarkably enough, is automatically
guaranteed in the framework of string theories \cite{cmmu}. On the other hand,
the existence of a non-renormalizable term as that of
eq.~(\ref{lamWo}) is also quite natural in superstring theories (see e.g.
ref.~\cite{cve}). This is equivalent
for practical purposes to the presence of a term $\lambda H_1H_2+$
h.c. in the K\"ahler potential, which resembles very much the
mechanism proposed in ref.~\cite{giuma} to solve the $\mu$ problem, namely the
presence of a term like $\lambda Z^*H_1H_2$ in $K$, where $Z^*$ is a
``hidden" field acquiring a large VEV.
%We will turn to this mechanism shortly.

\item[{\em b)}]

\hspace{10mm} In ref.~\cite{giurou} it was made the observation that in the
framework of any SUSY--GUT theory, starting again with $\mu=0$, an
effective $\mu$-term is generated by the integration  of the heavy
degrees of freedom. Assuming again a scenario with minimal K\"ahler
metric, so that the kinetic terms are canonical and the soft breaking
terms universal, the prediction for $B$ is once more $B=2m$, as in
eq.~(\ref{ABstr}).

\item[{\em c)}]

\hspace{10mm} Finally, let us note that the assumption of minimal
K\"ahler potential, $K$, is not really justified from string
theories. The form of $K$ in a large class of phenomenologically
interesting constructions was obtained in ref.~\cite{kahler} and the
corresponding soft breaking terms in
refs.~\cite{iblu,ccm}. It was noted,
however, that the assumption of universality can still be quite
reasonable (apart from phenomenologically desirable). It occurs
e.g. in the so--called dilaton--dominance limit \cite{kaplu93,bim}.
Concerning the
$\mu$ problem and the value of $B$, it was noted in ref.~\cite{bim}
that in the case of dilaton--dominance and a $\mu$--term arising as
suggested in ref.~\cite{giuma} or in ref.~\cite{cmmu} with
$\lambda=$ const. (see eq.~(\ref{lamWo})), then once again the value
$B=2m$ is obtained for large--size Calabi--Yau manifolds and orbifold
compactifications. In the latter case the two Higgses must belong to
the untwisted sector or possess similar modular weights.

\end{description}

\vspace{0.2cm}
\noindent It is not clear by now what is the reason behind the
coincidence in the prediction for $B$ in the previous three different
supergravity scenarios. Probably, it is a more general fact than just
a characteristic of these scenarios. In any case, we will refer (any
of) them as SUGRA II. The corresponding MSSM emerging from them is as
in minimal supergravity except for the value of $B$, which is given
by (\ref{ABstr}) instead of (\ref{ABmin}).

\section{Our calculation}

The aim of our calculation is to obtain the restrictions on the MSSM
from the $b\rightarrow s,\gamma$ decay in the two supergravity
scenarios, SUGRA I and SUGRA II, above defined. The scheme of our
procedure is the following. We start with the MSSM with initial
parameters
\be
\label{inPar}
\alpha_X,M_X,h_t,h_b,h_{\tau},\mu,m,M,A,B
\ee
By consideration of one of the two previous SUGRA theories we
eliminate the $B$ parameter through eq.~(\ref{ABmin}) or
eq.~(\ref{ABstr}). Then we demand consistency with all the
experimental data (apart from $b\rightarrow s,\gamma$). More
precisely, we require

\begin{description}

\item[{\em i)}] Correct unification of the gauge coupling
constants\footnote{Strictly speaking, this is not an experimental
observation, but it is a fact strongly suggested by the data that
nicely fits with the theoretical expectatives. We will turn to this
point below. }.

\item[{\em ii)}] Correct masses for all the observed particles.

\item[{\em iii)}] Masses for the unobserved particles compatible with
the experimental bounds.

\item[{\em iv)}] Correct electroweak breaking, i.e. $M_Z=M_Z^{exp}$.

\end{description}

\noindent Conditions {\em (i)--(iii)} allow to eliminate
$\alpha_X,M_X,h_t,h_b,h_{\tau}$, while {\em (iv)} allows to eliminate
one of the remaining parameters, which we choose to be  $\mu$. This
finally leaves three independent parameters, namely
\be
\label{inPar2}
m,M,A \; \; .
\ee
Actually, for many choices of these parameters there is no value of
$\mu$ capable of producing the
correct electroweak breaking. Therefore, this requirement restricts
the parameter space substantially. On the other hand, there can be two
branches of solutions depending on the sign of $\mu$. Finally, we
evaluate BR($b\rightarrow s,\gamma$) and
compare it with the experimental measurements, eventually obtaining
the restrictions on the initial parameters of theory coming from the
$b\rightarrow s,\gamma$ decay. We will also express the restrictions
in terms of some significative low--energy parameters, in particular
$\tan \beta$.

In order to be consistent, there is a number of refinements that have
to be considered in the calculation. Before explaining them, let us
stress that their importance is enhanced by the following fact. It is
known that in order to get a correct electroweak breaking in the
context of the MSSM a certain amount of fine--tuning between the
initial parameters is normally necessary \cite{barbi88}. It is a matter of
opinion what is the maximal acceptable amount of fine--tuning and how
to measure it \cite{barbi88,decar93,ross92,kane94}. But, in any case,
this means that rather small
variations on the initial parameters may have important implications
at low energy. Therefore, the determination of what are the initial
parameters compatible with the experimental conditions {\em
(i)--(iv)} has to be made in a very careful way.

The first refinement concerns condition {\em (i)}. Precision measurements
at LEP give a strong support \cite{amald91} to the expectation of
supersymmetric
unification of the gauge coupling constants\footnote{This unification
does not necessarily require a GUT. In particular, in superstring theories
all the gauge couplings are essentially the same at tree level \cite{witten},
even in the absence of a grand unification group. }
at $M_X\sim 10^{16}$ GeV, $\alpha_X\sim 0.04$. Actually, the
unification can only be achieved when the renormalization group
equations (RGE) of the gauge couplings are taken at two--loop order.
For consistency, all the supersymmetric thresholds (and the top quark one)
have to be taken into account in the running in a separate way. This
amounts to a technical problem since the values of the supersymmetric
masses depend themselves on the initial parameters of the theory at
$M_X$. Therefore, for a given model, i.e. for a choice of the initial
parameters, one cannot know a priori whether the resulting values of
$\alpha_1(M_Z),\alpha_2(M_Z),\alpha_3(M_Z)$ will be consistent with
their experimental values, namely \cite{amald91,langa}
\bea
\alpha_1(M_Z)&=& 0.016887 \pm 0.000040
\nonumber \\
\alpha_2(M_Z)&=& 0.03322 \pm 0.00025 \label{alfexp} \\
\alpha_3(M_Z)&=& 0.124 \pm 0.006 \nonumber
\eea
This means that in order to include the perturbative unification in a
consistent way an iterative process is necessary to adjust the
initial parameters so that (\ref{alfexp}) is fulfilled.

A second refinement has to do with the masses of the fermions. The
Yukawa couplings of the top and bottom quarks and the tau lepton,
$h_t,h_b,h_\tau$, play an important role in the radiative electroweak
breaking. Their boundary conditions have to be chosen so that the
experimental masses are fitted according to condition {\em (ii)}.
However, the running masses defined as $m_t(Q)=\langle H_2\rangle
h_t(Q)$, $m_b(Q)=\langle H_1\rangle h_b(Q)$, $m_\tau(Q)=\langle H_1\rangle
h_\tau(Q)$ do not coincide with the physical (pole) masses
$M_t,M_b,M_\tau$. In particular, for the top quark \cite{gray}
\be
\label{mtphys}
M_t=\left\{1+{\displaystyle\frac{4}{3}}
{\displaystyle\frac{\alpha_S(M_t)}{\pi}}+
\left[16.11-1.04\sum_{i=1}^{5}\left(1-\frac{M_i}{M_t}\right)\right]
\left(\frac{\alpha_S(M_t)}{\pi}\right)^2
\right\} m_t(M_t).
\ee
where $M_i$, $i=1,\ldots,5$ represent the masses of the five lighter
quarks, and similar expressions for the other quarks. In our
calculation we have used the recent evidence for top quark production
at CDF with a mass $M_t=174\pm17$ GeV \cite{abe94}, taking the central value.
Notice, however, that the masses in the BR($b\rightarrow s,\gamma$)
expressions, eqs.~(\ref{amps}), are the running masses at the electroweak
scale. Incidentally, it is amusing to realize that $m_t(M_Z)$ is
extremely close to the pole mass $M_t$ (while this does not happen for the
other particles).

Our third refinement, and the most important one, has to do with the
electroweak breaking process. The vacuum expectation values (VEVs) of
the two Higgses, $v_1=\langle H_1 \rangle$, $v_2=\langle H_2 \rangle$
are to be obtained from the minimization of the Higgs potential. The
tree level part of this in the MSSM has the form
\be
\label{Vo}
V_o=m_1^2 |H_1|^2 + m_2^2 |H_2|^2 + 2\mu B H_1H_2
+ \frac{1}{8}(g^2+{g'}^2)(|H_2|^2-|H_1|^2)^2\;\;,
\ee
where all the parameters are understood to be running parameters
evaluated at the renormalization scale $Q$. By a suitable redefinition
of the phases of the fields it is always possible to impose $v_1,v_2>0$.
As was clarified by
Gamberini et al.~\cite{gambe90}, $V_o$ and the corresponding tree
level VEVs $v_1^o(Q),v_2^o(Q)$ are strongly $Q$--dependent. In order
to get a much more scale independent potential the one--loop
correction $\Delta V_1$ is needed. This is given by
\be
\label{DeltaV1}
\Delta V_1={\displaystyle\sum_{j}}{\displaystyle\frac{n_j}{64\pi^2}}
M_j^4\left[\log{\displaystyle\frac{M_j^2}{Q^2}}
-\frac{3}{2}\right]\;\;,
\ee
where $M_j^2(\phi,t)$ are the tree-level (field--dependent) mass
eigenstates and $n_j$ are spin factors. In this way, the minimization
of $V=V_o+\Delta V_1$
gives one-loop VEVs $v_1(Q),v_2(Q)$ much more stable against
variations of the $Q$ scale. In general, there is a region of $Q$
where $v_1(Q),v_2(Q)$ are remarkably $Q$--stable and a particular
scale, $\hat{Q}$, always belonging to that region, at which
$v_1(\hat{Q}),v_2(\hat{Q})$ essentially coincide with
$v_1^o(\hat{Q}),v_2^o(\hat{Q})$. This is illustrated in Fig.~1 with a
particular example. The fact that $\hat{Q}$ always belong to the
stability region is not surprising since at $\hat{Q}$ the one--loop
correction $\Delta V_1$ is necessarily small, which is the right
criterion to choose an appropriate renormalization scale that
minimizes the size of the potentially large logarithms, thus
optimizing the perturbative expansion. $\hat{Q}$ is a certain average
of the masses contributing to $\Delta V_1$ in eq.~(\ref{DeltaV1}).
When we move far from $\hat{Q}$ the logarithms become large and the
perturbative expansion breaks down. As a result, $v_1(Q),v_2(Q)$
become strongly $Q$--dependent and eventually meaningless.
Consequently, when $\hat{Q}$ is substantially larger than $M_Z$ (and
this is the typical case), it is a bad approximation to minimize
$V=V_o+\Delta V_1$ at $Q=M_Z$ as is very commonly done. This is again
illustrated in Fig.~1. In our calculation we have evaluated $v_1,v_2$
at $\hat{Q}$ and then obtained $v_1(Q),v_2(Q)$ at any scale by using
the renormalization group running of the $H_1$, $H_2$ wave functions.
This is necessary in order to get the various masses appearing in the
problem at the appropriate scale. Furthermore, we have included in
eq.~(\ref{DeltaV1}) {\em all} the supersymmetric spectrum since, as
was stressed in \cite{decar93}, some common approximations, such as
considering
only the top and stop contributions, can lead to wrong results.

\vspace{0.2cm}
\noindent The previous refinements imply a whole iterative process in
order to determine a particular set of sensible initial parameter
[eq.~(\ref{inPar})]. This is in fact the most painful task of the
entire calculation. Then, for each model it is quite straightforward
to obtain the mass matrices of the supersymmetric particles and thus
the mass eigenvalues, diagonalization matrices, etc, to be plugged in
the expressions for the branching ratio, eqs.~(\ref{amps}). The results and
their implications are presented and discussed in the next section.

\section{Results}

We will consider the cases SUGRA I and SUGRA II in a separate way.

\vspace{0.3cm}
\noindent {\bf SUGRA I}

\vspace{0.2cm}
\noindent As discussed in the previous section, the consideration of
the physical requirements {\em (i)--(iv)} in the supergravity theory
at hand restricts the number of independent parameters to the set
$m,M,A$ [see eq.~(\ref{inPar2})]. In order to present the results in a
comprehensible way, let us make for the moment the assumption $m=M$.
The corresponding plots of the branching ratio BR($b\rightarrow
s,\gamma$) (in short BR) versus the remaining parameter, $A/m$, for
different values of $m$ are given in Fig.~2a (branch $\mu>0$) and Fig.~2b
(branch $\mu<0$). The lowest represented values of $m$ correspond to
the current experimental lower limits on supersymmetric particles \cite{pdg}.
The SM prediction for the BR and the CLEO upper and lower bounds are also
shown. In both figures we observe that the
MSSM result smoothly approaches the SM one as $m$ becomes larger,
though the limit is achieved quite slowly. For $\mu>0$ this trend
is only observed for $m\geq 200$ GeV, since for $m\leq 200$ GeV
the BR behaves in a much more ``unpredictable" way.

It is clear from the figures that the present CLEO bounds
exclude an appreciable region of the parameter space in both cases
$\mu>0$ and  $\mu<0$. It is interesting to trade the high energy
parameter $A$ by the low energy parameter $\tan \beta=v_2/v_1$ in the
representation. In this way we obtain the Figs.~3a, 3b. The lowest
value of $\tan \beta$ in the two figures is $\tan \beta =2$, which
essentially corresponds to the top infrared-fixed-point
value\footnote{For $M_{top}=174$ GeV and $h_t$ with the infrared
fixed point boundary condition, $\tan \beta \sim 1.9$. Lower values of
$\tan \beta$ cannot be achieved within the MSSM with universal soft
terms.}. We can see from the figures that for $m\geq 200$ GeV and both
$\mu>0$, $\mu<0$, for
each value of $m$ there
is a maximum acceptable value for $\tan \beta$, e.g. for $m=300$ GeV
we obtain $\left.\tan \beta\right|_{max}=12$ ($\mu>0$), 17 ($\mu<0$).
In general for $m\geq 200$ GeV  the restrictions are stronger for positive
values of $\mu$ and small values of $m$. For $m\leq 200$ GeV, however,
the situation is different: whereas for $\mu<0$ the restrictions become
very strong (only a narrow range of $\tan \beta$ is allowed), for $\mu>0$
there appear large windows of allowed values of $\tan \beta$. Here the
CLEO lower bound plays a relevant role.

The pattern of the restrictions is perhaps better appreciated by
plotting the branching ratio vs. $m$ for different values of
$\tan\beta$, as is done in Figs.~4a, 4b. For $\mu>0$, Fig.~4a,
we see that for small values of $m$, there is an allowed window
that becomes wider as $\tan\beta$ decreases. For large values of $m$
the result approaches the SM one, eventually becoming compatible
with the experimental bounds. For $\tan\beta<3$ the whole range of
$m$ is allowed, a fact that can change as soon as the experimental
bounds become a bit better, especially the upper one. For $\mu<0$,
Fig.~4b, we see that the BR behaves in a much more monotonic way. In
particular, there are no low-energy windows and there is a one-to-one
relation between the minimum acceptable values of $m$ and $\tan\beta$.
It is clear from the figures that the trend to the SM result is so slow
that an improvement of the CLEO bounds (particularly the upper one) on
BR of order $10^{-4}$ would push the lower limits
on $m$ to the TeV region (except for the above--mentioned windows in the
$\mu>0$ case), conflicting with the fine-tuning \cite{barbi88,decar93} and
cosmological \cite{kane94} constraints.
Strictly speaking the previous results apply only to the $m=M$ case,
but their sensitivity to departures from this assumption is rather
small, as illustrated in Fig.~5.

We would like at this point to compare the so far presented results
with some of the pre-existing ones in the literature
\cite{marco94,berto94,barge94,goto94}. It is in
fact worth-noticing that the results of some of these references are
rather incompatible each other (see e.g. refs.~\cite{marco94,goto94})
although the
comparison is in general very difficult to perform. This comes from
the fact that the parameters used to exhibit the dependence of the
branching ratio are very assorted (usually they are low energy
parameters). On the other hand, in some papers it is not specified in
the plots what are the values of the non-plotted parameters. In any
case, it is possible to detect some disagreements between our
results and those of
ref.~\cite{marco94} and, especially,
ref.~\cite{goto94}\footnote{The origin of these
discrepancies is probably due to the refinements explained in sect.~3.
One of us (B.C.) is indebted to M.A. D\'{\i}az for his help to show
that this is essentialy the case in ref.~\cite{marco94}.}. In
the latter
it is claimed that for $\mu>0$ (with our convention of signs) the
$b\rightarrow s,\gamma$ constraints are dramatically strong, while
for $\mu<0$ they are almost irrelevant. Although we find a trend in
this direction, our constraints are not that extreme in any of both
senses (see Figs.~2-4). This can also be applied to the results of
ref.~\cite{berto94}, though the comparison is
difficult for the above mentioned reasons.
Finally, ref.~\cite{barge94} deals with the infrared fixed point scenario,
which in our case corresponds to $\tan\beta\simeq 2$. Here we find a
good agreement between our results and theirs.

\vspace{0.3cm}
\noindent {\bf SUGRA II}

\vspace{0.2cm}
\noindent We present the results for the SUGRA II scenario in a way
completely analogous to those of SUGRA I. In particular, Figs.~6, 7, 8
correspond to Figs.~2, 3, 4 respectively. We note here that the sign of
$\mu$ is always negative since (adopting the convention $v_1,v_2 > 0$)
for $\mu>0$ and $B=2m$ the necessary
electroweak breaking cannot be achieved. The SUGRA II results present
a similar pattern to those of SUGRA I with $\mu<0$.
Again, we find that for a given value of $m$ there is a maximum
acceptable value for $\tan \beta$, e.g. for $m=300$ GeV, $\left.\tan
\beta\right|_{max}=12$. The bound becomes less stringent as $m$
increases. As in the SUGRA I scenario, the SM
limit is very smoothly achieved as $m$ grows, as can be seen from
Fig.~8. Therefore, once again, an improvement of the CLEO bounds
(especially the upper one) of order $10^{-4}$ would amount to a
dramatical improvement of the MSSM constraints, pushing the limits on
$m$ to the TeV region and thus conflicting with fine-tuning and
cosmological constraints.

\section{Conclusions}
We have evaluated the branching ratio BR($b\rightarrow s,\gamma$) in
the MSSM that emerges from two attractive supergravity scenarios,
SUGRA I and SUGRA II defined in sect.~2, determining the corresponding
phenomenological restrictions on the independent parameters of the theory.
For consistency, we have
included in the calculation some one--loop refinements that have a
substantial impact on the results. Our numerical results show some
disagreements with part of the previous results in the literature, while they
are in agreement with others. For SUGRA I (minimal supergravity) we
find that the CLEO upper and lower bounds are capable to put efficient
restrictions on
$m,M$ (the universal scalar and gaugino masses) in both cases $\mu<0$
and $\mu>0$, though the latter tends to be more
restrictive. For the SUGRA II scenario the results are similar,
but here the relevant role is played by the CLEO upper bound. Since
for both scenarios the trend to the SM limit as $m$ increases is very
slow, an eventual improvement of the CLEO upper and lower bounds of
order $10^{-4}$ would strengthen dramatically the previously found
constraints. This would be enough to discard the SUGRA I
with $\mu<0$ and SUGRA II scenarios if no discrepancy is found with
the SM prediction,
while for SUGRA I with $\mu>0$ there will remain low-energy windows.

\section*{Acknowledgements}

We thank L. Ib\'a\~nez and J. Moreno for very useful discussions. We
also thank P. Bourdon, O. Diego and J.R. Espinosa for their invaluable
help with the computer.

\newpage

\pagestyle{empty}
\renewcommand{\topfraction}{1.}

%%%%%%%%%%%%%%%%%%%%%%%%FIGURA%%%%%%%%%%%%%%%%%%%%%%%%
\begin{figure}
%\psdraft
\centerline{
\psfig{figure=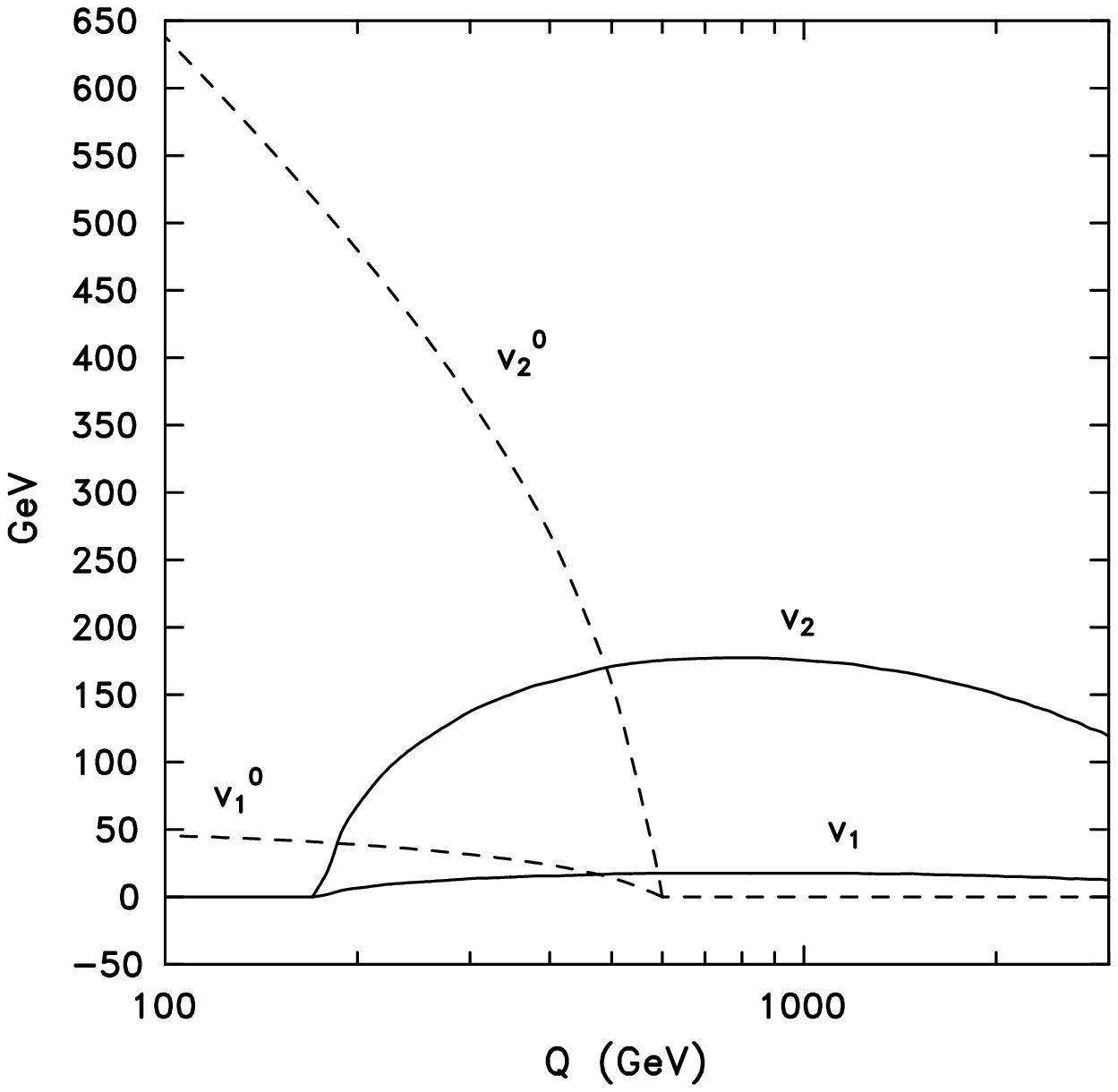,height=11cm,bbllx=6.cm,bblly=2.cm,bburx=15.cm,bbury=13cm}}
\caption{\tenrm\baselineskip=10pt Plot of
$v_1\equiv\langle H_1\rangle$, $v_2\equiv\langle H_2\rangle$
versus the $Q$ scale between 100 GeV and 3 TeV for the supersymmetric
model defined by $m=M=300$ GeV,
$A=317$ GeV, $B=A-m$, $\mu=403.76$ GeV, $h_t=0.568$, $h_b=0.063$,
$h_\tau=0.072$, $\alpha_X=0.0404$ (all
quantities defined at $M_X=1.60\times 10^{16}$ GeV).
Solid lines: complete one--loop results,
dashed lines: (renormalization improved) tree level results.
}
\end{figure}
%%%%%%%%%%%%%%%%%%%%%%%%%%%%%%%%%%%%%%%%%%%%%%%%%%%%%%
\begin{figure}
%\psdraft
\centerline{\vbox{
\psfig{figure=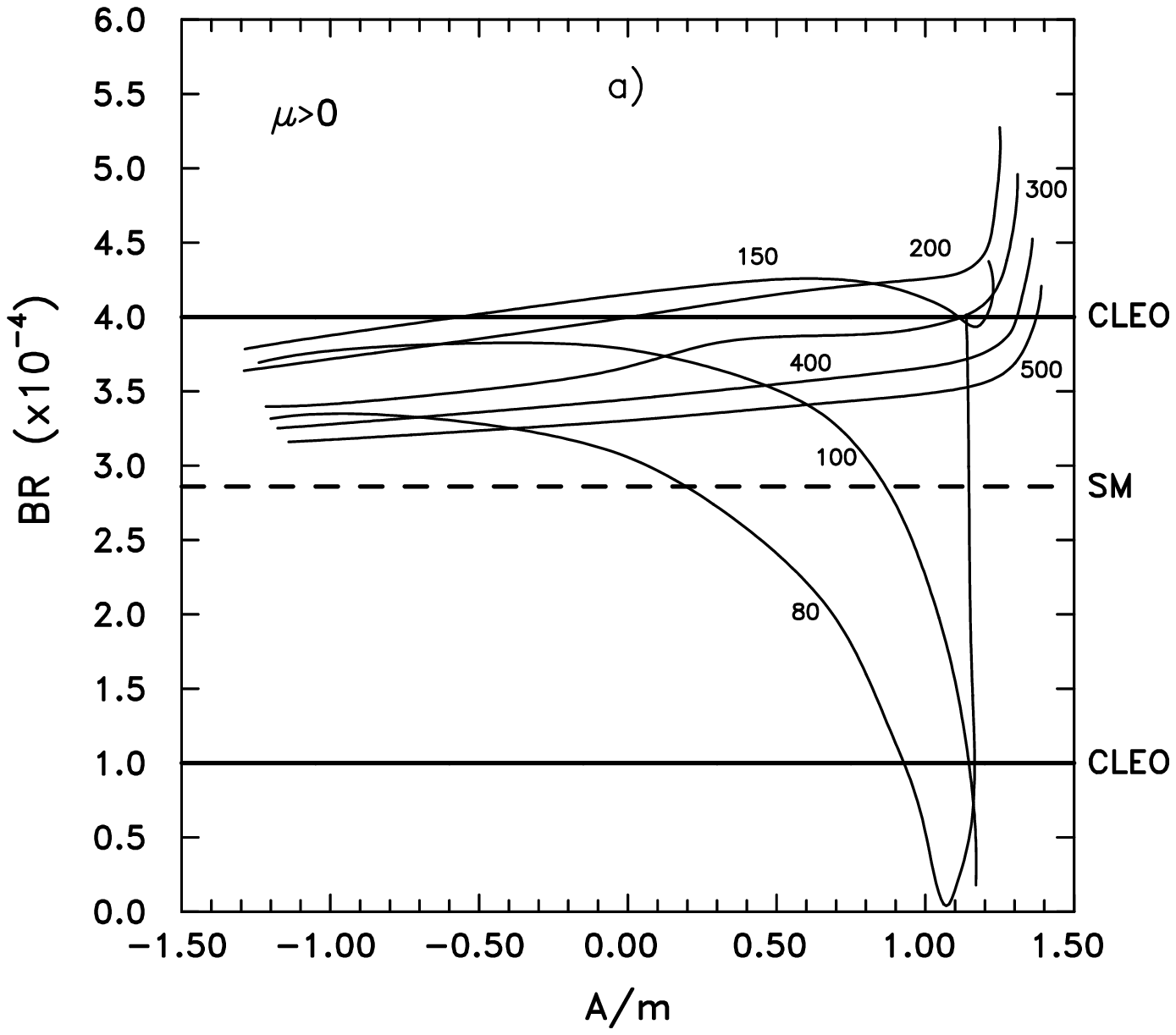,height=10cm,bbllx=6.cm,bblly=2.cm,bburx=15cm,bbury=15cm}
\psfig{figure=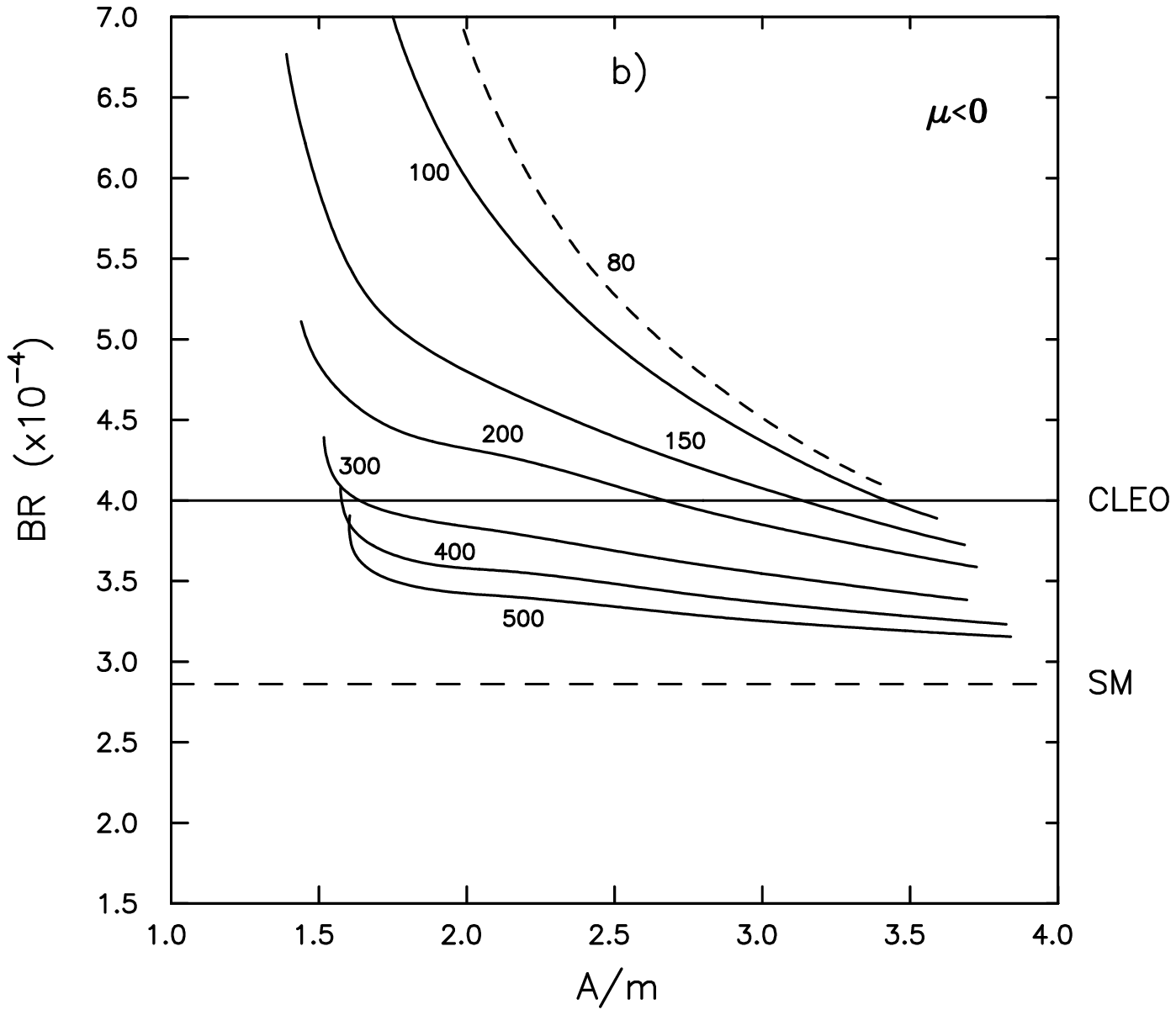,height=10cm,bbllx=6.cm,bblly=2.cm,bburx=15cm,bbury=15cm}
}}
\caption{\tenrm\baselineskip=10pt Plot of the branching ratio
BR($b\rightarrow s,\gamma$) (denoted BR) versus
$A/m$ for the SUGRA I scenario (minimal supergravity) with $m=M$ for
different
values of $m$ (namely, $m=$ 80, 100, 150, 200, 300, 400, 500 GeV)
: a) branch $\mu>0$; b) branch $\mu<0$. The dashed curves indicate that
the model becomes incompatible with the experimental lower bounds on
supersymmetric particles.
The Standard Model
prediction (SM) and the CLEO bounds are also shown in the figure.
}
\end{figure}
%%%%%%%%%%%%%%%%%%%%%%%%%%%%%%%%%%%%%%%%%%%%%%%%%%%%%%
\begin{figure}
%\psdraft
\centerline{\vbox{
\psfig{figure=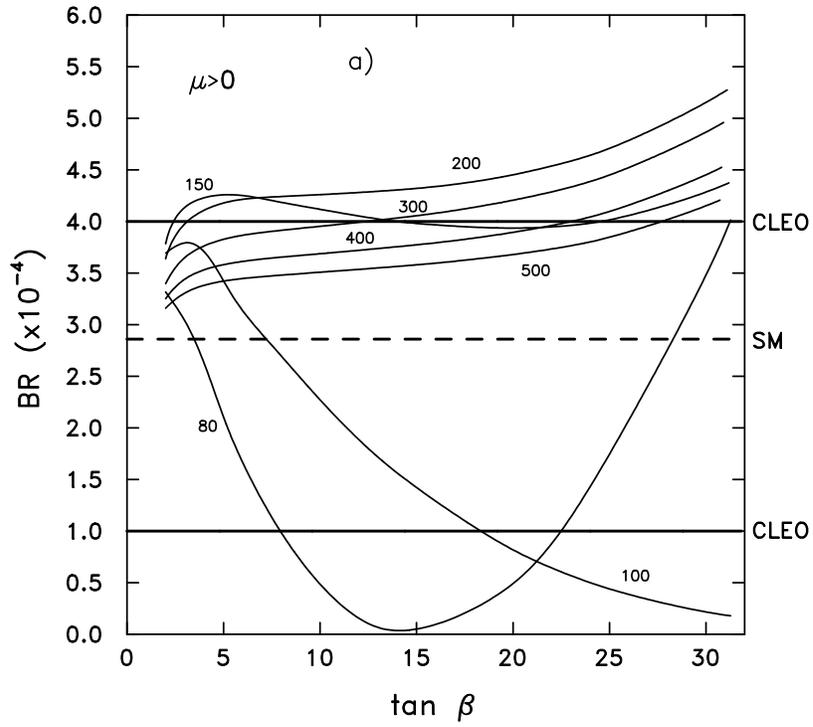,height=10cm,bbllx=6.cm,bblly=2.cm,bburx=15cm,bbury=15cm}
\psfig{figure=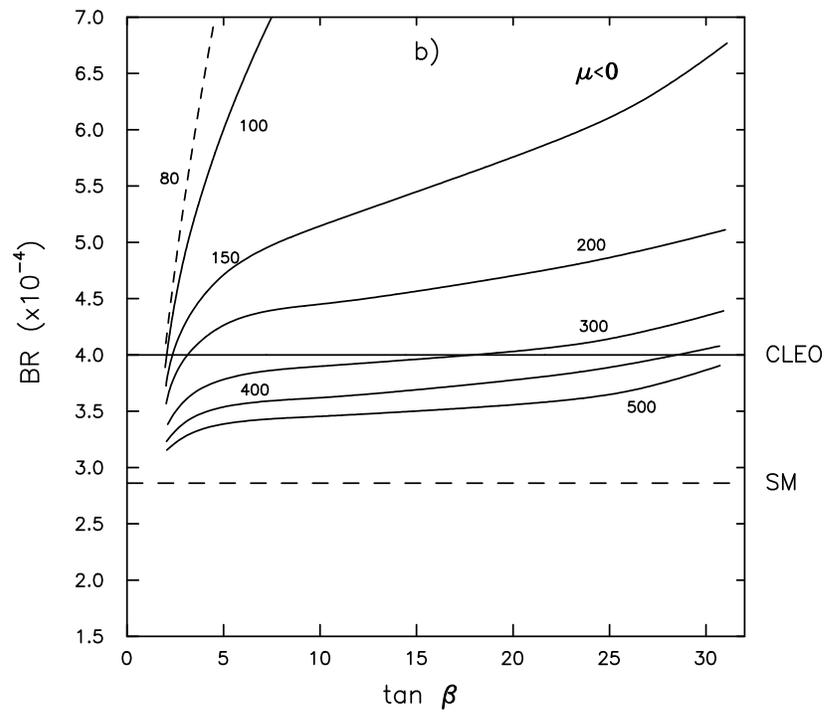,height=10cm,bbllx=6.cm,bblly=2.cm,bburx=15cm,bbury=15cm}
}}
\caption{\tenrm\baselineskip=10pt Plot of the branching ratio
BR($b\rightarrow s,\gamma$) versus $\tan \beta$ for the same models
and with the same conventions as in Fig.~2.
}
\end{figure}
%%%%%%%%%%%%%%%%%%%%%%%%%%%%%%%%%%%%%%%%%%%%%%%%%%%%%%
\begin{figure}
%\psdraft
\centerline{\vbox{
\psfig{figure=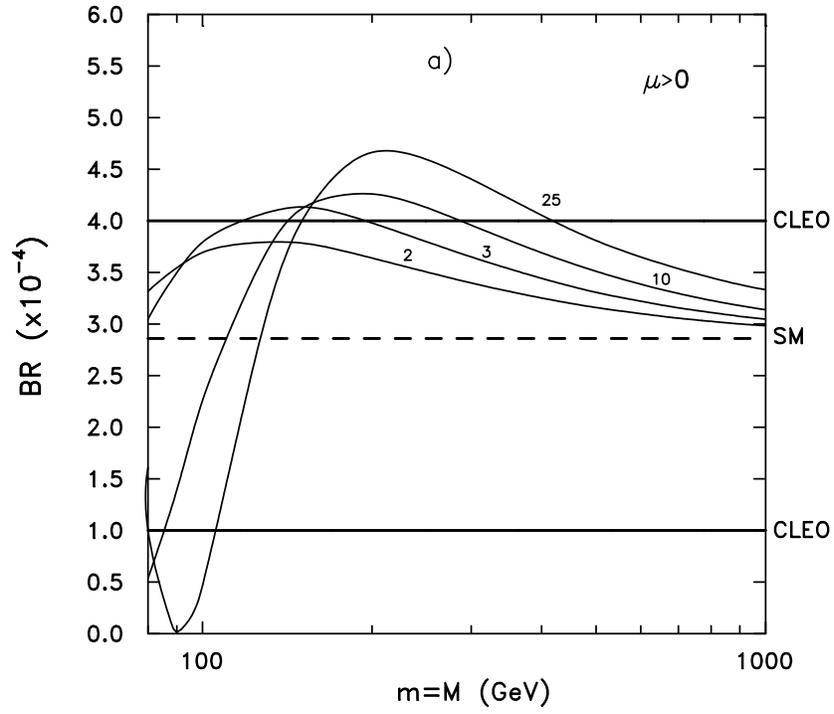,height=10cm,bbllx=6.cm,bblly=2.cm,bburx=15cm,bbury=15cm}
\psfig{figure=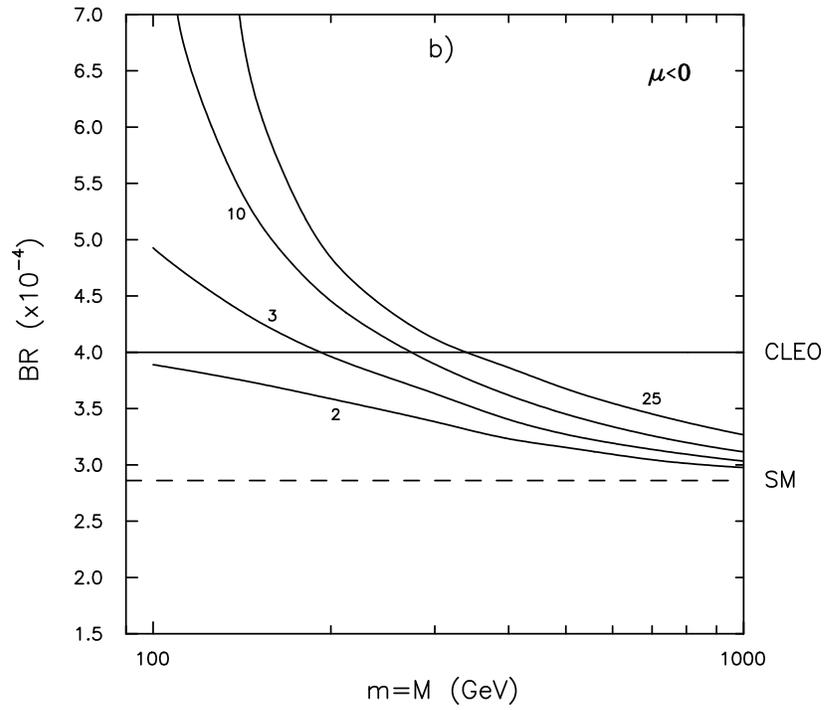,height=10cm,bbllx=6.cm,bblly=2.cm,bburx=15cm,bbury=15cm}
}}
\caption{\tenrm\baselineskip=10pt Plot of the branching ratio
BR($b\rightarrow s,\gamma$) versus
$m=M$ for different values of $\tan \beta$ (namely, $\tan \beta=$
2, 3, 10, 25) in the SUGRA I scenario:
a) branch $\mu>0$; b) branch $\mu<0$.
}
\end{figure}
%%%%%%%%%%%%%%%%%%%%%%%%%%%%%%%%%%%%%%%%%%%%%%%%%%%%%%
\begin{figure}
%\psdraft
\centerline{\vbox{
\psfig{figure=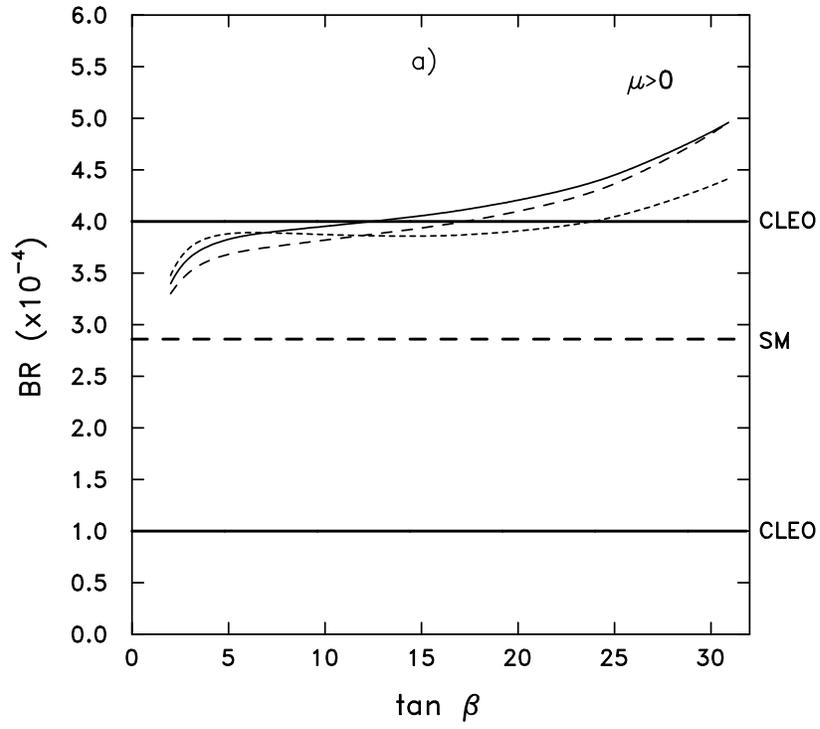,height=10cm,bbllx=6.cm,bblly=2.cm,bburx=15cm,bbury=15cm}
\psfig{figure=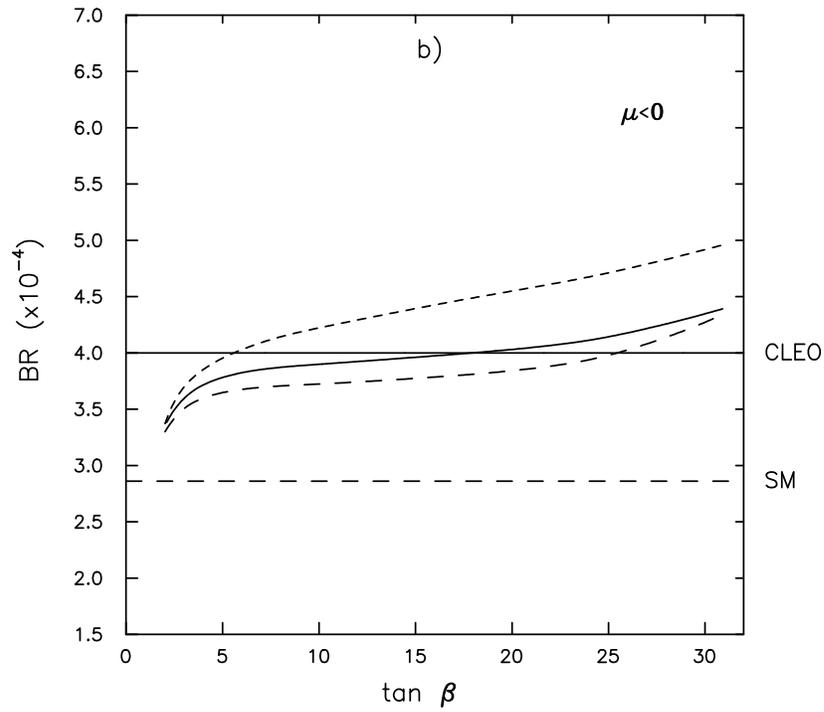,height=10cm,bbllx=6.cm,bblly=2.cm,bburx=15cm,bbury=15cm}
}}
\caption{\tenrm\baselineskip=10pt The same as in Fig.~3, but for
$m=300$ GeV and three different values
of $M$. Solid line: $M=300$ GeV, long-dashed line: $M=400$ GeV,
short-dashed line: $M=200$ GeV.
}
\end{figure}
%%%%%%%%%%%%%%%%%%%%%%%%%%%%%%%%%%%%%%%%%%%%%%%%%%%%%%
%%%%%%%%%%%%%%%%%%%%%%%%FIGURA%%%%%%%%%%%%%%%%%%%%%%%%
\begin{figure}
%\psdraft
\centerline{
\psfig{figure=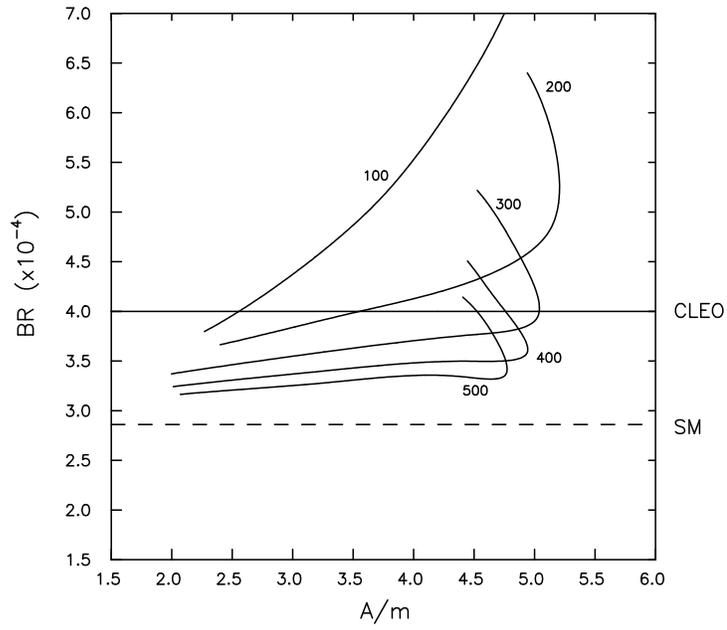,height=9.5cm,bbllx=6.cm,bblly=2.cm,bburx=15cm,bbury=16cm}}
\caption{\tenrm\baselineskip=10pt The same as Fig.~2, but for the SUGRA II
scenario (see definition in
the text), where only the $\mu<0$ branch
is meaningful.
}
\end{figure}
%%%%%%%%%%%%%%%%%%%%%%%%%%%%%%%%%%%%%%%%%%%%%%%%%%%%%%
%%%%%%%%%%%%%%%%%%%%%%%%FIGURA%%%%%%%%%%%%%%%%%%%%%%%%
\begin{figure}
%\psdraft
\centerline{
\psfig{figure=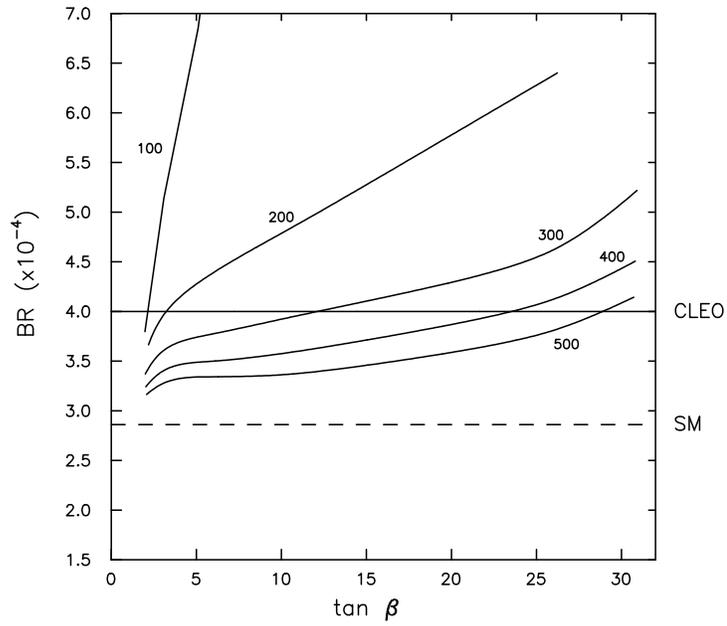,height=9.5cm,bbllx=6.cm,bblly=2.cm,bburx=15cm,bbury=16cm}}
\caption{\tenrm\baselineskip=10pt  The same as Fig.~3, but for the SUGRA II
scenario.
}
\end{figure}
%%%%%%%%%%%%%%%%%%%%%%%%%%%%%%%%%%%%%%%%%%%%%%%%%%%%%%
%%%%%%%%%%%%%%%%%%%%%%%%FIGURA%%%%%%%%%%%%%%%%%%%%%%%%
\begin{figure}
%\psdraft
\centerline{
\psfig{figure=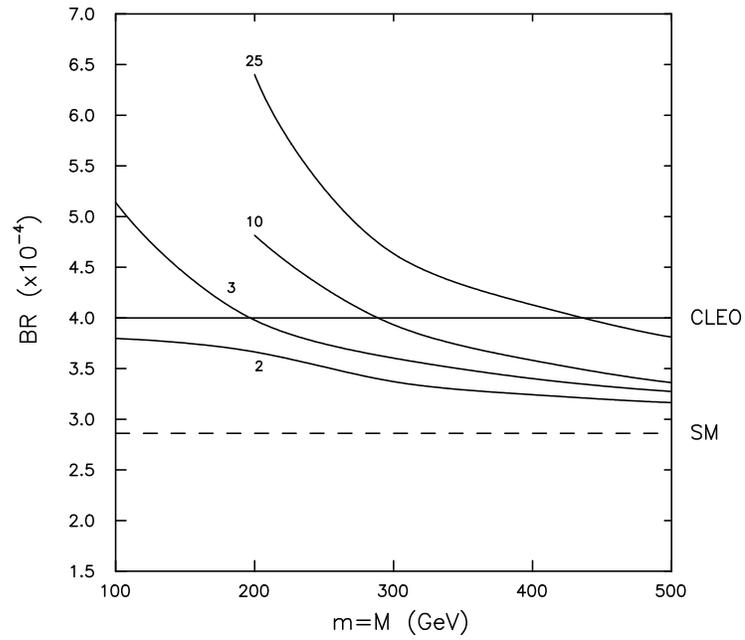,height=9cm,bbllx=6.cm,bblly=2.cm,bburx=15cm,bbury=15cm}}
\caption{\tenrm\baselineskip=10pt The same as Fig.~4, but
for the SUGRA II scenario. The $\tan \beta =$ 10, 25 curves have
been cut where they start to conflict with the experimental lower
limits on the supersymmetric particles.
}
\end{figure}
%%%%%%%%%%%%%%%%%%%%%%%%%%%%%%%%%%%%%%%%%%%%%%%%%%%%%%

\end{document}